\documentclass[showpacs,aps,twocolumn]{revtex4}
\usepackage{amsmath}
\usepackage{amsfonts}
\usepackage{amssymb}
\usepackage{amsthm}
\usepackage{graphicx}
\usepackage{color}

\begin{document}


\title{Domain wall motion of magnetic nanowires under a static field} 
\author{X. R. Wang}
\affiliation{Physics Department, The Hong Kong University of 
Science and Technology, Clear Water Bay, Hong Kong SAR, China}
\author{P. Yan }
\affiliation{Physics Department, The Hong Kong University of 
Science and Technology, Clear Water Bay, Hong Kong SAR, China}
\date{\today}

\begin{abstract}
The propagation of a head-to-head magnetic domain-wall (DW) or a 
tail-to-tail DW in a magnetic nanowire under a static field along 
the wire axis is studied. Relationship between the DW velocity 
and DW structure is obtained from the energy consideration. 
The role of the energy dissipation in the field-driven DW motion 
is clarified. Namely, a field can only drive a domain-wall 
propagating along the field direction through the mediation of a 
damping. Without the damping, DW cannot propagate along the wire. 
Contrary to the common wisdom, DW velocity is, in general, 
proportional to the energy dissipation rate, and one needs to find 
a way to enhance the energy dissipation in order to increase the 
propagation speed. The theory provides also a nature explanation 
of the wire-width dependence of the DW velocity and velocity 
oscillation beyond Walker breakdown field. 
\end{abstract}
\maketitle
Manipulation of domain wall (DW) of magnetic nanowires by a 
field\cite{Ono,Cowburn,Erskine,Parkin1,Erskine1} and/or by a 
current\cite{Erskine1,Ber,Thia,Mae,Klaui,Parkin2} has recently 
attracted much attention because of its academic interest and 
potential applications in information storage\cite{Parkin}. 
Many interesting phenomena were discovered with limited 
understandings. For a tail-to-tail DW or a head-to-head DW as 
shown in Fig. 1 in a magnetic nanowire with its easy-axis 
along the wire axis, the DW will propagate in the wire under 
an external magnetic field parallel to the wire axis or a 
current through the wire. The propagating speed $v$ of the DW 
depends on the field strength\cite{Erskine,Parkin1} and/or the 
current density\cite{Erskine1,Parkin2}. For the field-driven 
DW motion, there exists a so-called Walker's breakdown field 
$H_W$\cite{Walker}. The DW velocity $v$ is proportional to 
the external field $H$ for $H<H_W$ and $H\gg H_W$. DW velocity 
decreases as the field increases between the two linear 
H-dependence regimes. The linear regimes are characterized 
by the DW mobility $\mu\equiv v/H$. For $H\gg H_W$, the DW 
velocity may also oscillate with time\cite{Walker,Erskine}. 
Experiments also found that the DW velocity is sensitive to 
both DW structures and wire width\cite{Ono,Cowburn,Erskine}.
Magnetization dynamics is governed by a nonlinear differential 
equation, called Landau-Lifschitz-Gilbert (LLG) equation. 
Spatial homogeneous solutions of this equation related to the 
optimal magnetization reversal of Stoner particles may be solved 
analytically\cite{xrw}, but general solutions for the DW motion 
are not known. Our current theoretical knowledge about DW 
motion is either from the seminar work of Walker\cite{Walker} 
or from the Slonczewski's one-dimensional model\cite{Slon} in 
which a DW is treated as a rigid body characterized by DW 
position and a cant angle from the demagnetic field of a film. 
Both theories are developed for one-dimensional transverse 
DW of a uniaxial magnetic wire with a demagnetic field. 
\begin{figure}[htbp]
 \begin{center}
\includegraphics[width=7.cm, height=4.cm]{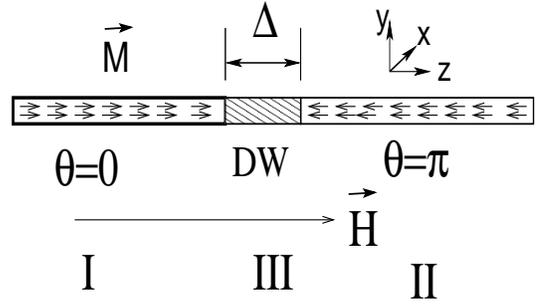}
 \end{center}
\caption{\label{fig1} Schematic diagram of a head-to-head 
domain wall (DW) in a magnetic nanowire. There are three 
phases in the wire, two magnetic domains I and II and DW of 
width $\Delta$. Domain I has its magnetization pointing to 
+z-direction ($\theta=0$) and the magnetization of Domain II 
is along -z-direction ($\theta=\pi$). III is the DW region 
whose magnetization structure could be very complicate. 
An external field is assumed to be along +z-direction. 
DW propagates along the field direction.}
\end{figure}

Though much progress have been made on both experimental and 
theoretical aspects, DW motion is still poorly understood. 
Slonczewski model is a great simplification of DW motion from 
the LLG equation. This model does not contain the detail 
structure of DW with DW width as an input parameter. 
Thus the theory can explain neither DW velocity oscillation 
nor many other interesting issues. Walker's theory is based 
on a particular solution of the LLG equation in a special 
case that does not distinguish different types of DWs. 
It predicts that the DW mobility is proportional to the DW 
width and inversely proportional to the damping constant, $\mu 
=\gamma\Delta/\alpha$, where $\gamma$ is gyromagnetic ratio, 
$\Delta$ is the DW width and $\alpha$ is damping constant. 
Walker's analytical analysis is useful in understanding some 
aspects of experimental findings, but it does not provide 
deep insight into DW motion such as the origin of the motion. 
In fact, Walker's expression of the DW mobility is strange 
because it is singular in $\alpha$. It does not tell us what 
will happen when damping goes to zero (the limit of $\lim
_{H\rightarrow 0}\lim_{\alpha\rightarrow 0}$ does not exist, 
and the limit of $\lim_{\alpha\rightarrow 0}\lim_{H\equiv\beta 
H_W\rightarrow 0}$ depends on $\beta$ that is a numeric number 
smaller than 1). Although Walker's solution is only correct 
for a strictly one-dimensional uniaxial wire with a demagnetic 
field and there is no proof of its correctness for other 
types of wires, it was widely used to understand DW motion. 
Thus, it is important to prove whether $\mu=\gamma\Delta
/\alpha$ is true for an arbitrary magnetic wire and 
to know how the result should be modified if it is not. 

In this letter, we shall examine the field-induced DW propagation 
from an angle different from those of Walker and Slonczewski. 
Firstly, we shall, from a simple model calculation, show that 
no static head-to-head DWs or tail-to-tail DWs are allowed in 
the presence of an external magnetic field along the wire axis. 
Secondly, energy conservation principle prevents a DW from 
propagating in the absence of a damping even under a static 
field, and it leads to a general relationship between the DW 
velocity and DW structure. Contrary to the common wisdom, DW 
velocity is proportional to the energy dissipation rate. 
Thus, the intrinsic critical field for DW propagation in a 
wire is zero, and the DW velocity is in general proportional to 
damping constant. Finally, our theory can reproduce the results 
of both Walker's and Slonczewski's models. Furthermore, the 
present theory attributes a DW velocity oscillation for $H\gg 
H_W$ to the periodic motion of the DW, either the precession 
of the DW or oscillation of the DW width. 

In a magnetic material, magnetic domains are formed in order to 
minimize the stray field energy. A DW that separates two domains 
is defined by the balance between the exchange energy and the 
magnetic anisotropy energy. The stray field plays little role 
in a DW structure. To describe a head-to-head DW in a magnetic 
nanowire, let us consider a wire with its easy-axis along the wire 
axis (the shape anisotropy dominates other magnetic anisotropies  
and makes the easy-axis along the wire when the wire is small 
enough) which is chosen as the z-axis as illustrated in Fig. 1. 
Since the magnitude of the magnetization $\vec M$ does not 
change in the LLG equation\cite{xrw}, the magnetic state of 
the wire can be conveniently described by the polar angle 
$\theta(\vec x,t)$ (angle between $\vec M$ and the z-axis) and 
the azimuthal angle $\phi(\vec x, t)$. The magnetization energy 
is mainly from the exchange energy and the magnetic anisotropy 
because the stray field energy is negligible in this case. 
The wire energy can be written in general as 
\begin{equation}
\begin{split}
& E=\int F(\theta,\phi,\vec\nabla\theta,\vec\nabla\phi)d^3\vec x,\\ 
& F=f(\theta,\phi)+J[(\vec\nabla\theta)^2+\cos^2\theta
(\vec\nabla\phi)^2]-MH\cos\theta, 
\end{split}\label{energy}
\end{equation}
where $f$ is the energy density due to all kinds of magnetic 
anisotropies which has two equal minima at $\theta=0$ and 
$\pi$ ($f(\theta=0,\phi)=f(\theta=\pi,\phi)$), $J-$term is 
the exchange energy, $M$ is the magnitude of magnetization, 
and $H$ is the external magnetic field along z-axis. 
In the absence of $H$, a head-to-head static DW that separates 
$\theta=0$ domain from $\theta=\pi$ domain (Fig. 1) can exist 
in the wire. The domain structure is determined by the equations 
\begin{equation}
\begin{split}
&\frac{\delta E} {\delta \theta}=2J\nabla^2\theta-\frac{\partial f}
{\partial \theta}+2J\sin\theta\cos\theta (\vec\nabla\phi)^2=0, \\
&\frac{\delta E} {\delta \phi}=2J\vec\nabla\cdot(\cos^2\theta \vec
\nabla\phi)-\frac{\partial f}{\partial \phi}=0. 
\end{split}\label{DW}
\end{equation}

In order to show that no intrinsic static head-to-head DW is allowed 
in the presence of an external field ($H\neq 0$), consider a special 
case that the wire is 1D and $f$ is a function of $\theta$ only. 
Then Eq. \eqref{DW} should include the external field and becomes  
\begin{equation}
\begin{split}
&2J\frac{d^2\theta}{dz^2}-\frac{\partial f}{\partial \theta}
-HM\sin\theta+
2J\sin\theta\cos\theta (\frac{d\phi}{dz})^2=0, \\
&2J\cos^2\theta \frac{d\phi}{dz}=const. 
\end{split}\label{DW1}
\end{equation} 
The second equation requires $d\phi/dz=0$. Since $\phi=const.$ 
and $\theta=0$ or $\pi$ on the both sides of the DW, $d\phi/dz=0$.
Thus, one can multiply the first equation by $d\theta/dz$ and the 
integration of the first equation yields $J(d\theta/dz)^2-
f(\theta)-HM\cos\theta=C=const$. However, for a head-to-head DW 
as shown in Fig. 1, $C=-f(0)-HM$ in the far left is not compatible
with $C=-f(\pi)+HM$ in the far right. Thus, Eq. \eqref{DW1} 
does not have a head-to-head DW solution for $H\neq 0$. 
In other words, a DW in a nanowire under an external field must be 
time dependent that could be either a local motion or a propagation 
along the wire. It should be clear that above argument is only true 
for an intrinsic head-to-head DW, and it shall fail with defect  
pinning that changes Eq. \eqref{DW1}. Indeed static DWs exist 
in the presence of a weak field in reality because of pinning. 

The following results are based on the fact that {\it a static 
field can be neither an energy source nor an energy sink}. 
Without a damping, the system is a {\it Hamiltonian system}, 
and will move on an equal energy contour by energy conservation. 
This viewpoint was important in understanding magnetization 
reversal of Stoner particles in a static field\cite{xrw1}, and 
its extension was used to develop new strategies\cite{xrw2} for 
the magnetization reversal of Stoner particles. From the LLG 
equation\cite{xrw1}, one can show that the energy damping rate is 
\begin{equation}\label{diss}
\frac{dE}{dt}=-\frac{\alpha\gamma}{1+\alpha^2}\int_{-\infty}
^{+\infty }\left(\vec M\times\vec H_{eff}\right)^2d^3\vec x, 
\end{equation}
where $\vec H_{eff}=-\frac{\delta F}{\delta\vec M}$ is the 
effective field. In regions I and II or inside a static DW, 
$\vec M$ is parallel to $\vec H_{eff}$. Thus no energy 
dissipation is possible there. The energy dissipation can 
only occur in the DW region when $\vec M$ is not parallel to 
$\vec H_{eff}$. 

The total energy of the wire equals the sum of the energies 
of regions I, II, and III, $E=\sum_{i=I,II,III}E_i$. 
$E_I$ increases while $E_{II}$ decreases when the DW propagate 
from left to the right along the wire. The net energy change 
$E_I+E_{II}$ in region I plus II due to the DW propagation is 
\begin{equation}\label{diss1}
\frac{d(E_{I}+E_{II})}{dt}=-2HMvA,
\end{equation}
where $v$ is the DW propagating speed, and $A$ is the cross 
section of the wire. The energy of region III should not change 
much because the DW width $\Delta$ is defined by the balance 
of exchange energy and magnetic anisotropy, and is usually order 
of $10\sim 100nm$. A DW cannot absorb or release too much energy, 
and can at most adjust temporarily energy dissipation rate. 
In other words, $\frac{dE_{III}}{dt}$ is either zero or fluctuates 
between positive and negative values with zero time-average. 
Since energy release from the magnetic wire should be equal to 
the energy dissipated (to the environment), one has 
\begin{equation}
-2HMvA +\frac{dE_{III}}{dt}=-\frac{\alpha\gamma}{1+\alpha^2}
\int_{III}\left(\vec M\times \vec H_{eff}\right)^{2}d^3\vec x. 
\end{equation}
Thus, the relationship between the DW velocity and the DW structure is 
\begin{equation}\label{main}
v= \frac{\alpha\gamma M}{2(1+\alpha^2)HA}\int_{III}\left(\vec m\times 
\vec H_{eff}\right)^{2}d^3\vec x +\frac{1}{2HMA}\frac{dE_{III}}{dt},
\end{equation}
where $\vec m$ is the unit vector of $\vec M$. Eq. \eqref{main} is 
our central result. Obviously, the right side of this equation 
is fully determined by the DW structure. Time averaged velocity is 
\begin{equation}\label{ave}
\bar v= \frac{\alpha\gamma M}{2(1+\alpha^2)HA}\int_{III}
\overline{\left({\vec m\times \vec H_{eff}}\right)^{2}}d^3\vec x, 
\end{equation}
where bar denotes time average. 

Because there is no intrinsic static head-to-head DWs in the 
presence of a static field, the first term in the right side 
of Eq. \eqref{main} will be positive and non-zero since a 
time dependent DW requires $\vec m\times \vec H_{eff}\neq 0$. 
Thus time-averaged $v$ is always positive. It implies {\it 
zero intrinsic critical-field for DW propagation} which is 
consistent with Walker's result\cite{Walker}. A DW can have 
two possible types of motion under an external magnetic field. 
One is that a DW behaves like a {\it rigid body} propagating  
along the wire. This is the basic assumption in Slonczewski 
model\cite{Slon,Thia} and Walker's solution for $H<H_W$. 
Obviously, DW energy is time-independent, $\frac{dE_{III}}{dt}=0$. 
The other one is that the DW structure varies with time while it 
propagates in the wire. It means that the snap-shots of the DW at 
different time are different. In the first type of motion, if the 
DW keep its static structure in the absence of the external field, 
then the first term in the right side of Eq. \eqref{main} shall 
be proportional to $a\Delta A H^2$, where $a$, characterize the DW 
structure, is a numerical number of order of 1 (smaller than 1). 
This is because the effective field due to the exchange energy 
and magnetic anisotropy is parallel to $\vec M$, and does not 
contribute to the energy dissipation. Thus, in this case, 
$v=a\frac{\alpha\gamma \Delta}{1+\alpha^2}H$ with $\mu=a\frac
{\alpha\gamma \Delta}{1+\alpha^2}$. To show that $a$ depends on 
material parameters, let us consider the Walker's\cite{Walker} 
model in which $f=-\frac{K_1}{2}M^2\cos^2\theta+ \frac{K_2}{2} 
M^2 \sin^2\theta\cos^2\phi,$ here $K_1$ and $K_2$ measure the 
magnetic anisotropies along the easy and hard axes, respectively. 
From Walker's trial function 
\begin{equation}
\begin{split}
& \ln\tan\frac{\theta(z,t)}{2}=c(t)\left[z-\int_0^tv(\tau)
d\tau \right], \\
&\phi (z,t)=\phi (t), 
\end{split}
\end{equation}
where $c^{-1}(t)$ is the DW width $\Delta$, 
Eq. \eqref{diss} becomes 
\begin{equation}\label{walker1}
\frac{dE}{dt}=-\frac{2\alpha\gamma A}{c(1+\alpha ^{2})}
\left[K_2^2M^{3}\sin^2\phi\cos^2\phi +H^2M\right],  
\end{equation}
and DW energy change rate is 
\begin{equation}\label{walker2}
\frac{dE_{III}}{dt}=\frac{d}{dt}\int_{III} F(\theta,\phi,
\vec\nabla\theta,\vec\nabla\phi)d^3\vec x =4JA\cdot \dot c(t). 
\end{equation}
Substituting Eqs. \eqref{walker1} and \eqref{walker2} into 
Eq. \eqref{main}, and together with Walker's conjectures of 
$C$ and $\phi$, one can easily reproduce Walker's DW velocity 
expression for both $H< H_W$ and $\gg H_W$. For example, 
for $H< H_W=\alpha K_2M/2$, $c^{-1}(t)=\Delta=const.$ and 
Eq. \eqref{main} gives 
\begin{equation}
v=\frac{\alpha\Delta\gamma}{1+\alpha^2}\left[1+
\left(\frac{K_2M\sin\phi\cos\phi}{H}\right)^2\right]H.
\end{equation}
This velocity expression is the same as that of the 
Slonczewski model\cite{Slon,Thia} in a one-dimensional wire. 
In Walker's analysis, $\phi$ is fixed by $K_2$ and $H$ 
through $K_2M\sin\phi\cos\phi=\frac{H}{\alpha }$. 
Using this $\phi$ in the above velocity expression, Walker's 
mobility coefficient $\mu=\frac{\gamma\Delta}{\alpha}$ 
is recovered. This inverse damping relation is from the 
particular potential landscape in $\phi$-direction. 
One should expect different result if the shape of the 
potential landscape is changed. Thus, this expression should 
not be used to extract the damping constant\cite{Ono,Erskine}.

A DW may be substantially distorted from its static structure 
when $H\gg H_W$ as it was revealed in Walker's analysis. 
Generally speaking, a physical system under a constant driving 
force will first try a fixed point solution\cite{xrw3}. 
It will go to other types of more complicated solutions only 
if a fixed point solution is not possible. In terms of DW 
motion, a domain wall will arrange itself as much as possible 
to satisfy Eq. (3). Of course, it is known that a head-to-head 
DW solution is only possible for Eq. (2) but not for Eq. (3). 
Thus, one should expect that $\vec M$ is almost parallel to $H
_{eff}$ except a small perpendicular component of order of $H$. 
One may assume $\vec M\times H_{eff}\simeq bMH\sin\varphi$ with 
$b$ much smaller than 1. This assumption agrees with the 
minimum energy dissipation principle\cite{Mea} since $|\vec M
\times H_{eff}|=MH\sin\theta$ when $\vec M$ for $H=0$ is used, 
and any modification of $\vec M$ should only make $|\vec M 
\times H_{eff}|$ smaller. The DW propagating speed is then 
$v=b'\alpha\gamma\Delta H/(1+\alpha^2)$, a linear relation in 
both $\Delta$ and $H$, but a smaller DW mobility because of 
the smallness of $b$. This smaller mobility at $H\gg H_W$ leads 
naturally to a negative differential mobility between $H<H_W$ 
and $H\gg H_W$! In other words, this negative differential 
mobility is due to the transition of the DW from a high energy 
dissipation structure to a lower one. This picture tells us 
that one should try to make a DW capable of dissipating as much 
energy as possible if one wants to achieve a high DW velocity. 
This is very different from what people would believe from 
Walker's special mobility formula of inverse proportion of the 
damping constant. To increase the energy dissipation, one may 
reduce defects and surface roughness other than increasing the 
damping constant. The reason is, by minimum energy dissipation 
principle, that defects are extra freedoms to lower $|\vec M
\times H_{eff}|$ because, in the worst case, defects will not 
change $|\vec M\times H_{eff}|$ when $\vec M$ without defects 
are used. 

In the case that the DW precess around wire axis or DW width 
breathes periodically or both motion are present, one should 
expect both $\frac{dE_{III}}{dt}$ and energy dissipation rate 
oscillate with time. According to Eq. \eqref{main}, DW velocity 
will also oscillate. DW velocity should oscillate periodically 
if only one type of DW motion (precession or DW breathing) 
presents, but it could be very irregular if both motions 
are present and the ratio of their periods is irrational. 
Indeed, this oscillation was observed in a recent 
experiment\cite{Erskine}. It should be pointed out that the DW 
precession motion was already detected experimentally by MOKE 
broadening technology\cite{Erskine}. How can one understand 
the wire-width dependence of the DW velocity? According to Eq. 
\eqref{main}, the velocity is a functional function of DW 
structure which is very sensitive to the wire width. 
For example, for a very narrow wire, only transverse DW is 
possible while a vortex DW is preferred for a very wide wire 
(large than DW width). Different vortices yield different 
values of $|\vec M\times H_{eff}|$, which in turn results in 
different DW propagation speed. 

The correctness of our central result Eq. \eqref{main} depends 
only on the LLG equation, the general energy expression of Eq. 
\eqref{energy}, and the fact that a static magnetic field can 
be neither an energy source nor an energy sink of a system. 
It does not depend on the details of a DW structure as long 
as the DW propagation is induced by a static magnetic field. 
In this sense, our result is very general and robust, and  
it is applicable to an arbitrary magnetic wire. 
However, it cannot be applied to a time-dependent field or 
the current-induced DW propagation, at least not directly.  
Also, it may be interesting to emphasize that there is no 
inertial in the DW motion within LLG description since this 
equation contains only the first order time derivative. 
Thus, there is no concept of mass in this formulation. 

In conclusion, a global view of the field-induced DW propagation 
is provided, and the importance of energy dissipation in the 
DW propagation is revealed. A general relationship between the 
DW velocity and the DW structure is obtained. The result says: 
{\it no damping, no DW propagation along a magnetic wire}. 
It is shown that the intrinsic critical field for a head-to-head 
DW is zero. This zero intrinsic critical field is related to 
the absence of a static head-to-head DW in a magnetic field 
parallel to the nanowire. Thus, a non-zero critical field can 
only come from the pinning of defects or surface roughness.  
The observed negative differential mobility is due to 
the transition of a DW from a high energy dissipation 
structure to a low energy dissipation structure. Furthermore, 
the DW velocity oscillation is attributed to either the DW 
precession around wire axis or from the DW width oscillation. 

This work is supported by Hong Kong UGC/CERG grants 
(\# 603007 and SBI07/08.SC09).

\end{document}